\documentclass[aps,pra,twocolumn,superscriptaddress,pdfLaTeX]{revtex4-2}

\usepackage{amsmath}
\usepackage{bm}
\usepackage{graphicx}
\usepackage{hyperref}
\usepackage{amsmath}
\usepackage{ulem}
\hypersetup{colorlinks=true,citecolor=magenta,linkcolor=magenta,urlcolor=magenta}

\begin{document}


\title{Geometrical scheduling of adiabatic control without information of energy spectra}


\author{Yuta Shingu}
\affiliation{Department  of  Physics, Graduate School of Science, Tokyo  University  of  Science,  Shinjuku,  Tokyo  162-8601,  Japan.}

\author{Takuya Hatomura}
\email[]{takuya.hatomura@ntt.com}
\affiliation{NTT Basic Research Laboratories \& NTT Research Center for Theoretical Quantum Information, NTT Corporation, Kanagawa 243-0198, Japan}


\date{\today}

\begin{abstract}
Adiabatic control is a fundamental technique for manipulating quantum systems, guided by the quantum adiabatic theorem, which ensures suppressed nonadiabatic transitions under slow parameter variations. Quantum annealing, a heuristic algorithm leveraging adiabatic control, seeks the ground states of Ising spin glass models and has drawn attention for addressing combinatorial optimization problems. However, exponentially small energy gaps in such models often necessitate impractically long runtime to satisfy the adiabatic condition. Despite this limitation, improving the quality of approximate solutions remains crucial for practical applications.
The quantum adiabatic brachistochrone provides a method to enhance adiabaticity by minimizing an action representing nonadiabaticity via the variational principle. While effective, its implementation requires detailed energy spectra, complicating its use in quantum annealing. Shortcuts to adiabaticity by counterdiabatic driving offer alternative approaches for accelerating adiabatic processes. However, the theory of shortcuts to adiabaticity often faces challenges such as nonlocal control requirements, high computational cost, and trade-offs between speed and energy efficiency.
In this work, we propose a novel quantum adiabatic brachistochrone protocol tailored for quantum annealing that eliminates the need for energy spectrum information. Our approach builds on advancements in counterdiabatic driving to design efficient parameter schedules. We demonstrate the effectiveness of our method through numerical simulations on the transverse-field Ising chain and axial next-nearest neighbor Ising models.
\end{abstract}


\maketitle


\section{Introduction}

Adiabatic control is one of the fundamental techniques for harnessing quantum systems. 
A guiding principle of adiabatic control is the adiabatic theorem of quantum mechanics~\cite{Kato1950,Albash2018}. 
The adiabatic theorem tells us that transitions between different energy levels, i.e., nonadiabatic transitions, do not take place when parameters of a system vary slowly compared with the size of energy gaps. 
Various quantum algorithms based on adiabatic control have been proposed~\cite{Kadowaki1998,Farhi2000,Roland2002}.

Quantum annealing is such a heuristic quantum algorithm for finding the ground state of Ising spin glass~\cite{Kadowaki1998,Albash2018,Hauke2020}. 
In quantum annealing, we first prepare the trivial ground state of a simple Hamiltonian, e.g., the transverse-field Hamiltonian, and adiabatically transform it into the nontrivial ground state of Ising spin glass. 
Quantum annealing has been paid much attention because solutions of combinatorial optimization problems can be embedded in the ground states of Ising spin glass models and many social issues can be formulated as combinatorial optimization problems~\cite{Lucas2014,Yarkoni2022}. 
Even though such Ising spin glass models often exhibit exponentially small energy gaps, which require exponentially long runtime for obtaining the exact solutions due to the adiabatic condition~\cite{Jansen2007}, approximate solutions of quantum annealing encoded in low energy states of Ising spin glass models can be useful enough for many social issues. 
Therefore, even if it is difficult to obtain the ground state, improving quality of approximate solutions is still an important subject.

Quantum adiabatic brachistochrone is one of the methods for improving adiabaticity~\cite{Rezakhani2009,Rezakhani2010}. 
In quantum adiabatic brachistochrone, we introduce an action representing nonadiabaticity and minimize it based on the variational principle. 
The resulting geodesic equation gives optimal parameter schedules of a given Hamiltonian. 
It has been reported that quantum adiabatic brachistochrone with various actions can improve adiabaticity~\cite{Rezakhani2009,Rezakhani2010,Takahashi2019,Hatomura2019a,Kazhybekova2022}. 
However, we in principle need information of energy spectra to construct appropriate actions, and thus it might be difficult to apply quantum adiabatic brachistochrone to quantum annealing.

Use of shortcuts to adiabaticity~\cite{Demirplak2003,Berry2009,Chen2010,Torrontegui2013,Guery-Odelin2019,Hatomura2024} is another approach for improving adiabaticity. 
Various methods have been proposed as shortcuts to adiabaticity. 
In counterdiabatic driving, we apply additional terms counteracting diabatic changes to a reference Hamiltonian~\cite{Demirplak2003,Berry2009}. 
However, there are several obstacles to theoretical construction and experimental realization. 
Exact construction of additional terms requires exponentially large computational cost in general. 
Application of counterdiabatic driving to many-body systems requires time-dependent control of many-body and nonlocal interactions~\cite{DelCampo2012}. 
Moreover, there is a tradeoff between speedup and energy cost~\cite{Santos2015,Campbell2017}. 
Therefore, we usually adopt some approximations and aim to achieve moderate speedup and moderate fidelity with reasonable energy cost. 
Several approximate methods have been proposed for applying counterdiabatic driving to quantum annealing~\cite{Sels2017,Hatomura2018b,Ozguler2018} and their performance was studied in various models~\cite{Hartmann2019a,Passarelli2020,Hatomura2020a,Hayasaka2023}.

The variational approach is a prominent method of approximate counterdiabatic driving~\cite{Sels2017}. 
In this method, we make an ansatz on the operator form of additional driving and determine the time-dependence of coefficients based on the variational principle. 
Remarkably, this calculation does not require information of energy spectra. 
As mentioned above, it has been used to improve performance of quantum annealing~\cite{Hartmann2019a,Passarelli2020}. 
Recently, the algebraic theory of variational counterdiabatic driving was developed~\cite{Hatomura2021} and a computationally efficient approach incorporating the Lanczos method with the Krylov subspace was proposed~\cite{Bhattacharjee2023,Takahashi2024}. 
Moreover, the theory of variational counterdiabatic driving was utilized for detection of quantum phase transitions~\cite{Hatomura2021,Kim2024} and quantum chaos~\cite{Pandey2020,Bhattacharjee2023}. 
It suggests that the theory of variational counterdiabatic driving gives some information on energy spectra, while calculation of variational counterdiabatic driving does not require information of energy spectra.

In this paper, keeping application to quantum annealing in mind, we develop a quantum adiabatic brachistochrone protocol which does not require information of energy spectra. 
Key ingredients of our method are based on recent development of shortcuts to adiabaticity by counterdiabatic driving. 
We demonstrate our method by using the transverse-field Ising chain with and without next-nearest neighbor terms. 
The rest of the present paper is constructed as follows. 
We summarize the background theory in Sec.~\ref{Sec.background} and introduce our protocol in Sec.~\ref{Sec.protocol}. 
In Sec.~\ref{Sec.benchmark}, we conduct a benchmark test of our protocol. 
Section~\ref{Sec.discussion} is devoted to discussion on interpretation of obtained results. 
We conclude the present paper in Sec.~\ref{Sec.conclusion}

\section{Theory}

\subsection{\label{Sec.background}Background}

We consider a quantum system described by a time-dependent Hamiltonian $\hat{H}(\lambda)$, where $\lambda=\lambda(t)$ is a time-dependent parameter. 
For simplicity, in this paper, we focus on a single time-dependent parameter, but our scheme can be expanded to multiple time-dependent parameters.
Time evolution of the system $|\Psi(t)\rangle$ is governed by the Schr\"odinger equation, $i(\partial/\partial t)|\Psi(t)\rangle=\hat{H}(\lambda)|\Psi(t)\rangle$. 
Here and hereafter, we set $\hbar=1$.

In quantum adiabatic brachistochrone~\cite{Rezakhani2009,Rezakhani2010}, we obtain the optimal parameter schedule $\lambda=\lambda(t)$ by solving the geodesic equation
\begin{equation}
\begin{aligned}
&\ddot{\lambda}+\Gamma(\lambda)\dot{\lambda}^2=0, \\
&\Gamma(\lambda)=\frac{1}{2g(\lambda)}\partial_{\lambda}g(\lambda), 
\end{aligned}
\label{Eq.geodesic}
\end{equation}
where the dot symbol represents time derivative. 
We can suppress nonadiabatic transitions when a metric $g(\lambda)$ appropriately represents nonadiabaticity. 
See, Appendix~\ref{Sec.QAB} for details.

Counterdiabatic driving is another strategy for suppressing nonadiabatic transitions~\cite{Demirplak2003,Berry2009}. 
In counterdiabatic driving, we introduce the counterdiabatic Hamiltonian
\begin{equation}
\begin{aligned}
&\hat{H}_\mathrm{cd}(t)=\dot{\lambda}\hat{\mathcal{A}}(\lambda), \\
&\hat{\mathcal{A}}(\lambda)=i\sum_{\substack{n,m \\ (n\neq m)}}|n(\lambda)\rangle\langle n(\lambda)|\partial_{\lambda}m(\lambda)\rangle\langle m(\lambda)|, 
\end{aligned}
\label{Eq.cdham}
\end{equation}
where $\hat{\mathcal{A}}(\lambda)$ is known as the adiabatic gauge potential~\cite{Kolodrubetz2017}. 
The counterdiabatic Hamiltonian has the ability to completely cancel out nonadiabatic transitions. 
See, Appendix~\ref{Sec.nonadiabaticity} for details.

In the Krylov approach of approximate counterdiabatic driving~\cite{Takahashi2024,Bhattacharjee2023}, we express approximate adiabatic gauge potential $\hat{\mathcal{A}}^\ast(\lambda)$ as
\begin{equation}
\hat{\mathcal{A}}^\ast(\lambda)=i\sum_{i}\alpha_i(\lambda)\hat{O}_{2i-1},
\label{Eq.Krylov.AGP}
\end{equation}
where $\{\hat{O}_i\}$ is the set of the basis operators generated by the Lanczos method
\begin{equation}
\begin{aligned}
&b_0(\lambda)\hat{O}_0=\partial_\lambda \hat{H}, \\
&b_1(\lambda)\hat{O}_1=\mathcal{L}\hat{O}_0, \\
&b_i(\lambda)\hat{O}_i=\mathcal{L}\hat{O}_{i-1}-b_{i-1}(\lambda)\hat{O}_{i-2} & (i \geq 2),
\end{aligned}
\label{Eq.Lanczos}
\end{equation}
where $\mathcal{L}$ is defined by $\mathcal{L}\bullet=[\hat{H}(\lambda),\bullet]$ and $b_i(\lambda)$ is the normalization factor for $\hat{O}_i$ with the (rescaled) Hilbert-Schmidt norm, i.e., $b_i(\lambda)$ is determined so that $\|\hat{O}_i\|_\mathrm{HS}=\sqrt{(1/D)\mathrm{Tr}\hat{O}_i^\dag\hat{O}_i}$=1. 
Here, $D$ is the dimension of the system. 
The odd basis operators $\{\hat{O}_{2i-1}\}$ can span the exact counterdiabatic Hamiltonian (\ref{Eq.cdham}) if we compute Eq.~\eqref{Eq.Lanczos} until $b_i(\lambda)=0$ (see, Appendix~\ref{Sec.AGP}). 
However, the number of basis operators $\{\hat{O}_{i}\}$ typically scales exponentially with system size. 
Therefore, we usually truncate the Lanczos method (\ref{Eq.Lanczos}) at the $2d_{A}$th basis with a certain integer $d_A$.
By incorporating this expression (\ref{Eq.Krylov.AGP}) to the theory of variational counterdiabatic driving, we obtain the following linear equation
\begin{widetext}
\begin{equation}
\begin{pmatrix}
b_1^2+b_2^2 & b_2b_3 & 0 \\
b_2b_3 & b_3^2+b_4^2 & b_4b_5 \\
0 & b_4b_5 & b_5^2+b_6^2 &  \\
 &  &  & \ddots \\
 &  &  &  & b_{2d_A-1}^2+b_{2d_A}^2 \\
\end{pmatrix}    
\begin{pmatrix}\alpha_1 \\ \alpha_2 \\ \vdots \\ \alpha_{d_A}\end{pmatrix}	
=\begin{pmatrix}-b_0b_1 \\ 0 \\ \vdots \\ 0
\end{pmatrix}.
\label{eq: linear}
\end{equation}
\end{widetext}
By solving this equation, we can determine the coefficients $\{\alpha_i(\lambda)\}$. 
See, Appendix~\ref{Sec.AGP} for details.

It is known that the counterdiabatic Hamiltonian (\ref{Eq.cdham}) includes geometrical information of energy spectra (see, Appendix~\ref{Sec.geometrical} for details). 
As expected, approximate counterdiabatic Hamiltonians also have partial information of energy spectra and can be used to detect quantum phase transitions~\cite{Hatomura2021,Kim2024} and quantum chaos~\cite{Pandey2020,Bhattacharjee2023}. 
Thus, we expect that the amplitude of an approximate counterdiabatic Hamiltonian $\hat{H}_\mathrm{cd}^\ast(t)=\dot{\lambda}\hat{\mathcal{A}}^\ast(\lambda)$, i.e., 
\begin{equation}
\begin{aligned}
&\|\hat{H}_\mathrm{cd}^\ast(t)\|_\mathrm{HS}=\dot{\lambda}\sqrt{g^\ast(\lambda)}, \\
&g^\ast(\lambda)=\sum_k\Big(\alpha_k(\lambda)\Big)^2, 
\end{aligned}
\label{Eq.cdham.metric}
\end{equation}
can be used as a measure of nonadiabaticity.

\subsection{\label{Sec.protocol}Our method}

Our protocol for parameter scheduling is as follows: 
\begin{enumerate}
\item Calculate the recurrence formula (\ref{Eq.Lanczos}) by using symbolic computation~\cite{Meurer2017} and determine the truncated set of the basis operators $\{\hat{O}_i\}$. 
\item Construct Eq. (\ref{eq: linear}) by using symbolic computation and solve it.
\item Solve the geodesic equation (\ref{Eq.geodesic}) with $g(\lambda)=g^\ast(\lambda)$ in Eq.~(\ref{Eq.cdham.metric}).
\end{enumerate}
We remark on each step below.

{\it Step 1}. 
In disordered systems, the number of the basis operators in the Lanczos method (\ref{Eq.Lanczos}) rapidly increases compared with the system size. 
In practice, we have to truncate most of the high-order basis operators. 
For the next step, the number of the basis operators should be at most polynomial against the system size.

{\it Step 2}. 
Standard approaches for solving a linear equation use the LU decomposition.
Those approaches only require $\mathcal{O}(n)$ computational cost against its dimension $n$ since the matrix in Eq.~\eqref{eq: linear} is tri-diagonalized. 
Here, the dimension $n$ is equal to the number of the basis operators in the previous step.

{\it Step 3}. 
The quantity $g^\ast(\lambda)$ in Eq.~(\ref{Eq.cdham.metric}) satisfies the definition of the metric. 
Indeed, its multi-parameter expression is actually positive, real, and symmetric. 
Moreover, the geodesic equation (\ref{Eq.geodesic}) is rewritten as
\begin{equation}
\dot{\lambda}=\frac{C}{\sqrt{g(\lambda)}},
\end{equation}
with a constant $C$~\cite{Rezakhani2010}, which can easily be solved by using, for example, the standard Runge-Kutta methods.

\section{\label{Sec.benchmark}Numerical benchmarking}

We apply our protocol to quantum annealing for evaluating its performance. 
First, we overview quantum annealing. 
We consider a system consisting of $L$ qubits. 
The Hamiltonian of quantum annealing is given by
\begin{equation}
\hat{H}(\lambda)=\lambda\hat{H}_P+(1-\lambda)\hat{H}_V,
\end{equation}
with the problem Hamiltonian
\begin{equation}
\hat{H}_P=-\sum_{\substack{i,j=1 \\ (i<j)}}^LJ_{ij}\hat{Z}_i\hat{Z}_j-\sum_{i=1}^Lh_i^z\hat{Z}_i,
\end{equation}
and the driver Hamiltonian
\begin{equation}
\hat{H}_V=-\sum_{i=1}^L\hat{X}_i, 
\end{equation}
where we express the Pauli matrices as $\hat{X}_i$, $\hat{Y}_i$, and $\hat{Z}_i$ ($i=1,2,\dots,L$). 
We change the time-dependent parameter $\lambda$ from 0 to 1 so that $\hat{H}(0)=\hat{H}_{\mathrm{V}}$ and $\hat{H}(1)=\hat{H}_{\mathrm{P}}$. 
The solution of a given problem is embedded in the constants $J_{ij}$ and $h_i^z$ of the problem Hamiltonian. 
By choosing the initial state as $|+\cdots+\rangle$, we can estimate the ground state and the ground-state energy of $\hat{H}_{\mathrm{P}}$ where $|+\rangle$ is the eigenstate of $\hat{X}$ with the eigenvalue $+1$.
The simplest parameter schedule is $\lambda=t/T$, where $T$ is the annealing time. 
According to the literature of quantum annealing, we assume that all the energy and time parameters are dimensionless.

\subsection{Example 1: Transverse Ising chain}

As the first example, we consider the one-dimensional transverse-field Ising model with the periodic boundary condition, i.e., $J_{i,i+1}=J$ and $J_{L,L+1}=J_{L1}=J$ (otherwise $J_{ij}=0$), $h_{i}^z=0$, $\{\hat{X}_{L+1},\hat{Y}_{L+1},\hat{Z}_{L+1}\}=\{\hat{X}_{1},\hat{Y}_{1},\hat{Z}_{1}\}$. 
We set $J=1$ for simplicity.

The Lanczos method (\ref{Eq.Lanczos}) and the translational symmetry of the system tell us that the basis operators spanning the exact counterdiabatic Hamiltonian is given by
\begin{equation}
\hat{O}_{2k-1}=\frac{1}{\sqrt{2L}}\sum_{i=1}^L(\hat{Y}_i\hat{X}_i^{(k-1)}\hat{Z}_{i+k}+\hat{Z}_i\hat{X}_i^{(k-1)}\hat{Y}_{i+k}),
\label{Eq.basis.trans}
\end{equation}
where $k=1,2,\dots,L-1$, and
\begin{equation}
\left\{
\begin{aligned}
&\hat{X}_i^{(k-1)}\equiv\prod_{j=1}^{k-1}\hat{X}_{i+j},\quad\text{for }k=2,3,\dots,L-1, \\
&\hat{X}_i^{(0)}\equiv\hat{1}. 
\end{aligned}
\right.
\end{equation}
Notably, even if we adopt all these basis operators, the number of the basis operators is $L-1$, i.e., linear against the system size. 
As the set of the basis operators for the approximate counterdiabatic Hamiltonian, we adopt the truncated set of the basis operators $\{\hat{O}_{2k-1}\}_{k=1,2,\dots,d_A}$ with $d_A\le L-1$.

\begin{figure}
\includegraphics[width=0.5\textwidth]{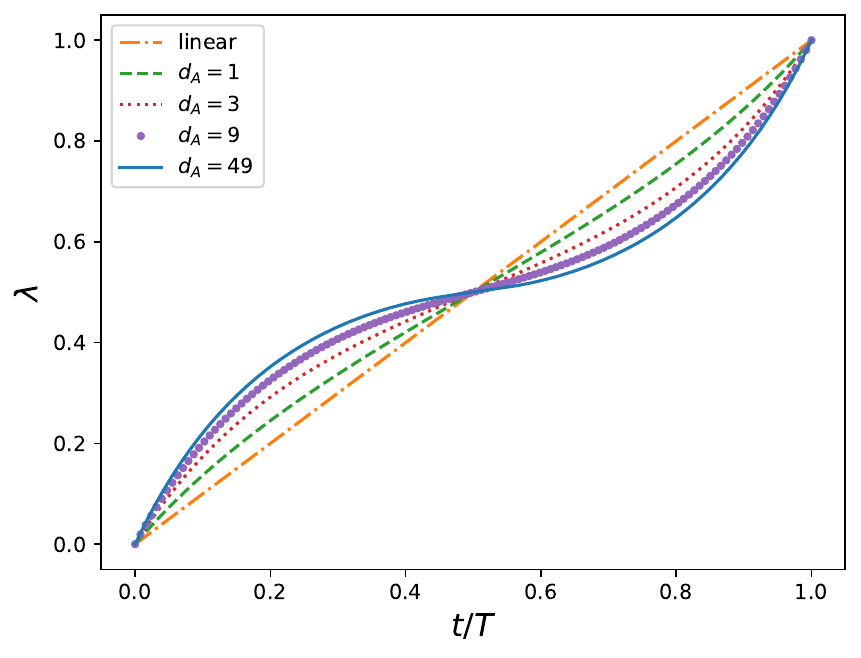}
\caption{\label{Fig.sche.Ising}Obtained schedules for the transverse-field Ising chain with $L=50$. The horizontal axis indicates the normalized time with the annealing time $T$. The solid line is computed with the full bases and the dash-dot line indicates the linear schedule. The other lines are obtained with the truncated method. }
\end{figure}

Figure~\ref{Fig.sche.Ising} shows the obtained schedules. 
The solid line is obtained by computing the full bases with $d_A=49$ ($L=50$) in Eq.~(\ref{Eq.Lanczos}) at each normalized time $t/T$. 
As a reference, we draw the linear schedule with the dash-dot line. 
The other line are obtained by truncating the Lanczos method with $d_A=1,3,9$. 
We find that the truncation method gives schedules similar to the full-bases one.

\begin{figure}
\includegraphics[width=0.5\textwidth]{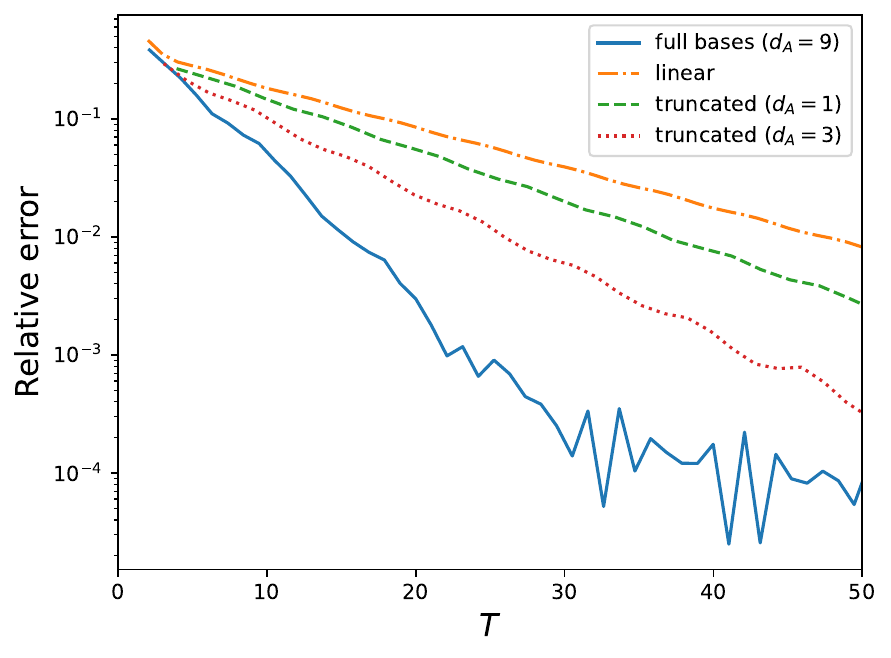}
\includegraphics[width=0.5\textwidth]{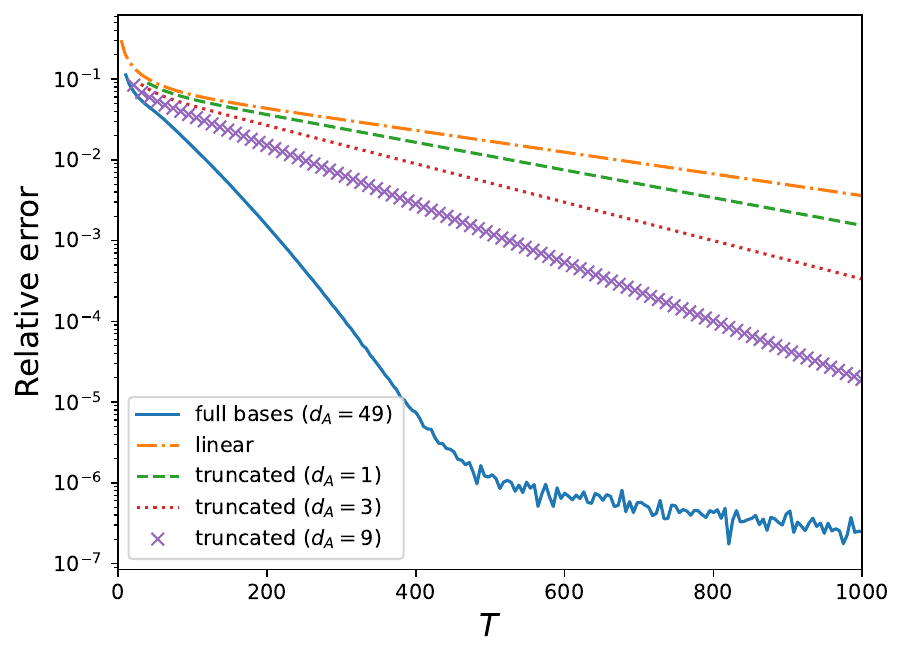}
\caption{\label{Fig.isingchainene}Relative error of quantum annealing in the transverse-field Ising chain with (top) $L=10$ and (bottom) $L=50$. The solid line is the result obtained with the full-bases method. The dash-dot line draws the performance of the linear schedule. The other lines indicate the result of the truncation method. All the schedules obtained with our strategy achieve lower relative errors than the linear schedule in all the annealing time $T$. }
\end{figure}

As a figure of merit, we consider the relative error, $|\langle\hat{H}_P\rangle-E_g|/|E_g|$, where $E_g$ is the ground-state energy. 
We plot the relative error in Fig.~\ref{Fig.isingchainene}. 
We find the scaling advantage of our method compared with the linear schedule.

\subsection{Example 2: Axial next-nearest neighbor Ising model}
We also compute numerical simulations for the axial next-nearest neighbor Ising (ANNNI) model~\cite{Selke1988}. The problem Hamiltonian with the periodic boundary is given by
\begin{equation}
\hat{H}_{\mathrm{P}}=-J\sum_{i=1}^L \hat{Z}_i \hat{Z}_{i+1} + k \sum_{i=1}^L \hat{Z}_i \hat{Z}_{i+2},
\end{equation}
which is known as a non-integrable extension of the Ising model.
Unlike the one-dimensional transverse-field Ising model, this model is complicated as it is difficult theoretically to obtain the basis operators.
Thus, we numerically compute the Lanczos method~\eqref{Eq.Lanczos}.
We set $J=1$ with a 6-qubits system in the numerical simulation.

\begin{figure}
  \includegraphics[width=0.5\textwidth]{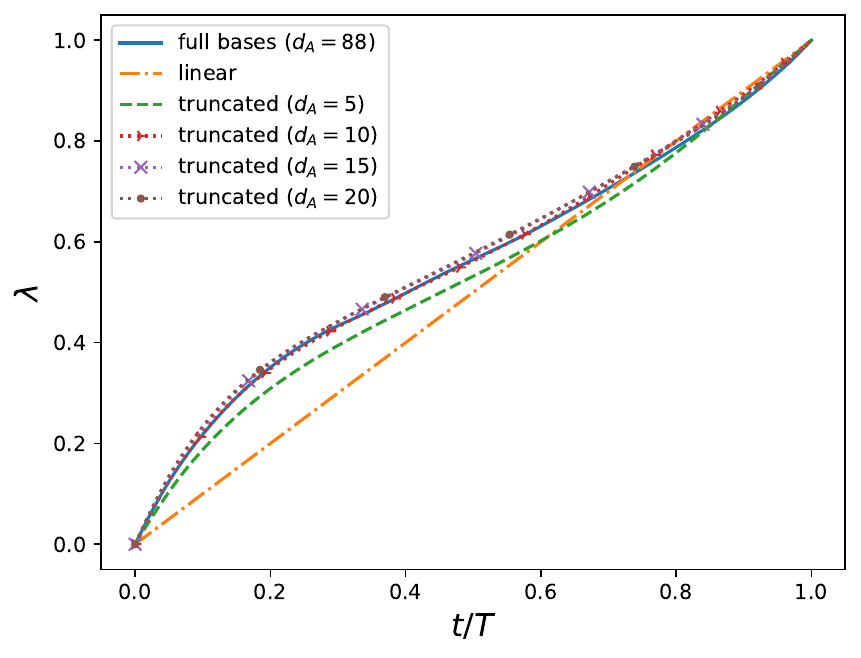}
  \caption{
  Obtained schedules for the ANNNI model in the ferromagnetic phase. The horizontal line indicates the normalized time with the annealing time $T$. The solid line is numerically computed with the full bases. The dash-dot line indicates the linear schedule. The other lines are obtained with the truncation method.
  }
  \label{fig: ANNNI model schedule}
\end{figure}
First, we set $k=0.3$, for which the ground state in a ferromagnetic state. 
Figure~\ref{fig: ANNNI model schedule} indicates obtained schedules with our method.
The blue solid line is obtained by computing the full bases with $d_A=88$ in Eq.~\eqref{Eq.Lanczos} at each normalized time $t/T$.
Here, $d_A=88$ is determined numerically rather than theoretically, as $b_{177}$ converges to 0 numerically.
We draw the dash-dot line as the linear schedule. 
The other lines are obtained by truncating the Lanczos method with $d_A=5,10,15,$ and 20.
Figure~\ref{fig: ANNNI model schedule} shows that the truncated lines are almost equal to the full-basis schedule even with  $d_A=10$.

\begin{figure}
  \includegraphics[width=0.5\textwidth]{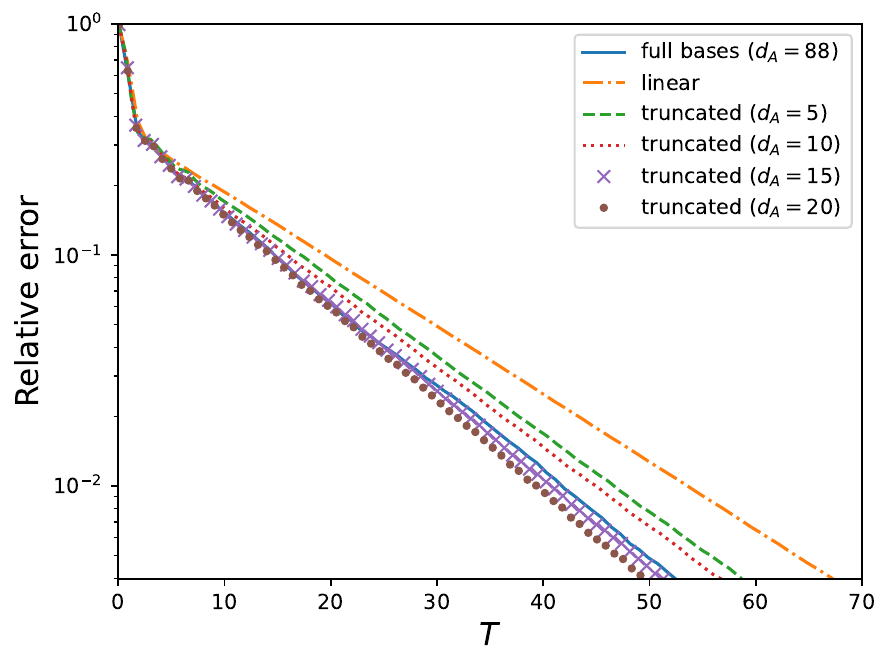}
  \caption{Relative error of quantum annealing in the ANNNI model (the ferromagnetic phase). 
  The horizontal line indicates the annealing time. 
  The solid line is the result obtained with the full-bases method.
  The dash-dot line draws the performance of the linear schedule.
  The other lines indicate the result of the truncation method.
  All the schedules obtained with our strategy achieve lower relative errors than the linear schedule in all the annealing time $T$.
  }
  \label{fig: ANNNI model result}
\end{figure}
In Fig.~\ref{fig: ANNNI model result}, we compare our method with the linear-schedule method by studying the performance of quantum annealing. The vertical line indicates the relative error for $\hat{H}_{\mathrm{P}}$ computed by $|\langle\hat{H}_{\mathrm{P}}\rangle - E_{\mathrm{g}}|/|E_{\mathrm{g}}|$ where $E_{\mathrm{g}}$ denotes the exact ground-state energy of $\hat{H}_{\mathrm{P}}$. Even with the 5-basis truncated schedule, we see a faster convergence than the linear schedule. 
Note that the truncation 20-basis schedule shows a slightly faster than the full-basis schedule.

\begin{figure}
  \includegraphics[width=0.5\textwidth]{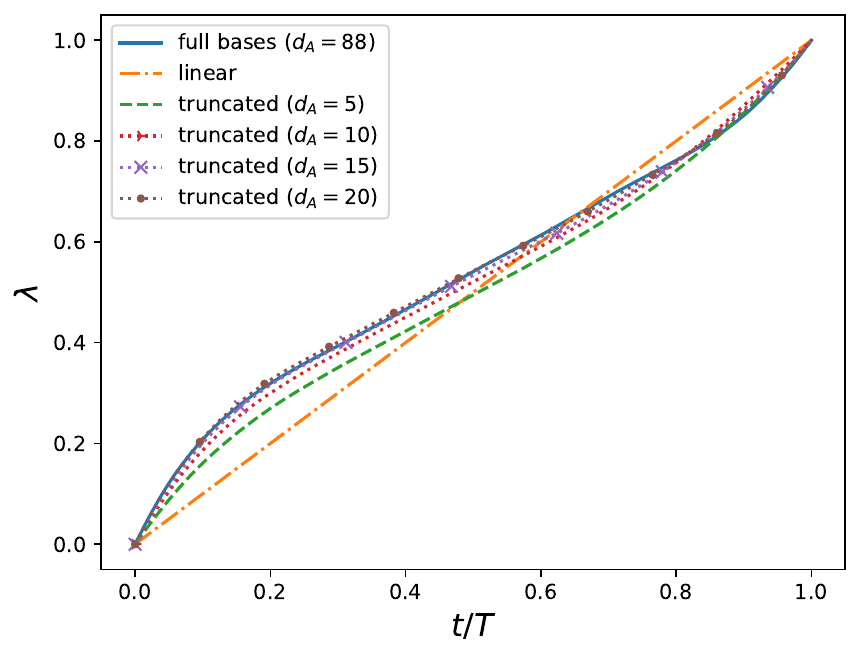}
  \caption{Obtained schedules for the ANNNI model in the antiferromagnetic-like phase. The horizontal line indicates the normalized time with the annealing time $T$. The solid line is numerically computed with the full bases. The dash-dot line indicates the linear schedule. The other lines are obtained with the truncation method.
  }
  \label{fig: ANNNI model schedule k=0.7}
\end{figure}

\begin{figure}
  \includegraphics[width=0.5\textwidth]{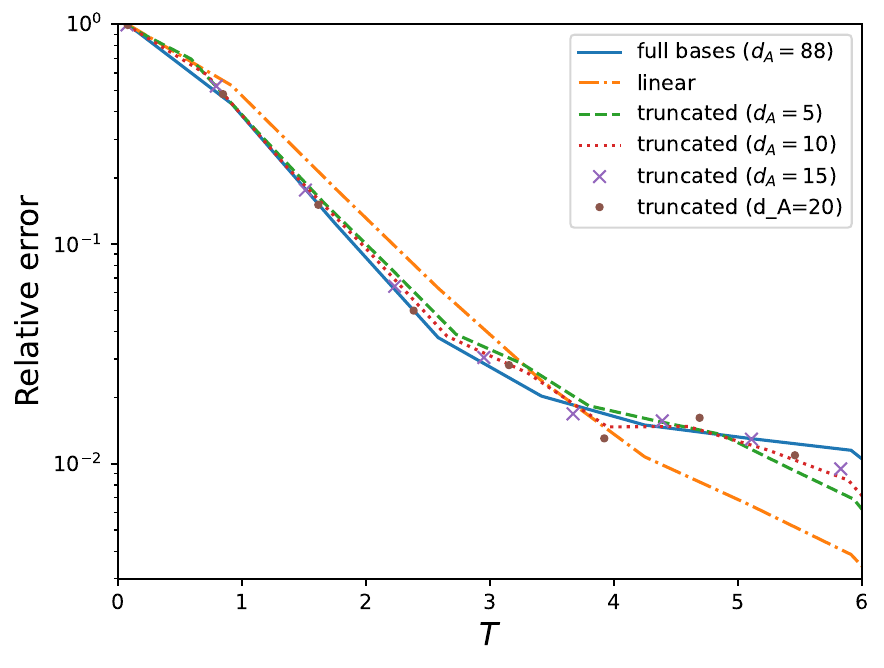}
  \caption{Relative error of quantum annealing in the ANNNI model (the antiferromagnetic-like phase). 
  The horizontal line indicates the annealing time. 
  The solid line is the result obtained with the full-bases method.
  The dash-dot line draws the performance of the linear schedule.
  The other lines indicate the result of the truncation method.
  Our schedules are better than the linear one until around $T=3.5$, while the relative error with the linear schedule shows lower relative errors than our method when $T$ is larger than around 3.5.
  }
  \label{fig: ANNNI model result k=0.7}
\end{figure}

We further compute numerical simulation with $k=0.7$, for which the ground state is an antiferromagnetic-like state $|\uparrow\uparrow\downarrow\downarrow\uparrow\uparrow\dots\rangle$.
Figure~\ref{fig: ANNNI model schedule k=0.7} draws the obtained schedules with our method. Figure~\ref{fig: ANNNI model result k=0.7} shows the performance with our schedules and the linear schedule.
We obtain the lower relative error than that with the linear schedule until around $T=3.5$, but the linear schedule is superior to our schedule when setting a larger annealing time. 
We will discuss this behavior in the next section. 

\section{\label{Sec.discussion}Discussion}

By applying our method to quantum annealing, we confirmed improvement in performance. 
Particularly, our strategy shows better results for the one-dimensional transverse-field Ising model and the ANNNI model in the ferromagnetic phase even when we truncate many terms from the exact adiabatic gauge potential. 
We also found that our schedule is superior to the linear one when we consider a nonadiabatic regime of the ANNNI model in the antiferromagnetic phase, while the linear schedule gives better results for the adiabatic regime.

For the ferromagnetic state of the ANNNI model, we observed non-monotonic improvement of performance against the number of truncation (Fig.~\ref{fig: ANNNI model result}) and found that the strength of the exact adiabatic gauge potential (\ref{Eq.strength.cdham}) is not always the optimal measure for nonadiabaticity. 
We interpret this result as a bad influence of redundant full spectral information included in the exact adiabatic gauge potential (\ref{Eq.cdham}), which may not be suitable for the ground-state search. 
We expect that such a bad influence increases in more complicated systems. 
Indeed, we observed more complicated behavior in the antiferromagnetic state of the ANNNI model (Fig.~\ref{fig: ANNNI model result k=0.7}), which has more geometric frustration than the ferromagnetic one and shows more complicated energy spectra~\cite{Selke1988}.

\begin{figure}
  \includegraphics[width=0.5\textwidth]{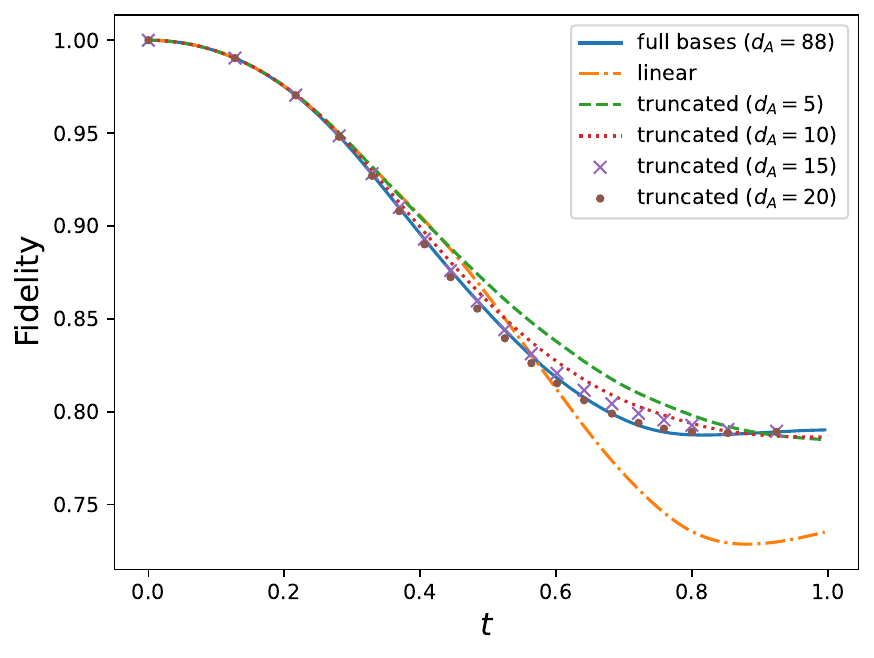}
  \caption{Fidelity of the obtained state to the ground state of the ANNNI model in the antiferromagnetic-like phase at each time $t$. 
  The annealing time $T$ is fixed as 1.75. 
  Our strategy finally achieves higher fidelity than the linear schedule.
  }
  \label{fig: fidelity_vs_t175}
\end{figure}

Now, we conduct a detailed analysis on the antiferromagnetic state of the ANNNI model. 
Let us consider the case setting a shorter annealing time than $T=3.5$. 
Figure~\ref{fig: fidelity_vs_t175} indicates the fidelity of the obtained state to the ground state, $|\langle\phi_{\mathrm{g}}(\lambda(t))|\Psi(t)\rangle|^2$, for the annealing time $T=1.75$, where $|\phi_{\mathrm{g}}(\lambda(t))\rangle$ is the ground state and $|\Psi(t)\rangle$ is the obtained state. 
The fidelity with each schedule is almost monotonically decreasing.
It means that non-adiabatic transitions occur throughout the dynamics and the systems are almost monotonically escaping from the ground state.
The fidelity with our schedules decreases to around 0.80, while the one with the linear schedule falls below 0.75.
Thus, our schedules effectively suppress non-adiabatic transitions compared with the linear one.

\begin{figure}
  \includegraphics[width=0.5\textwidth]{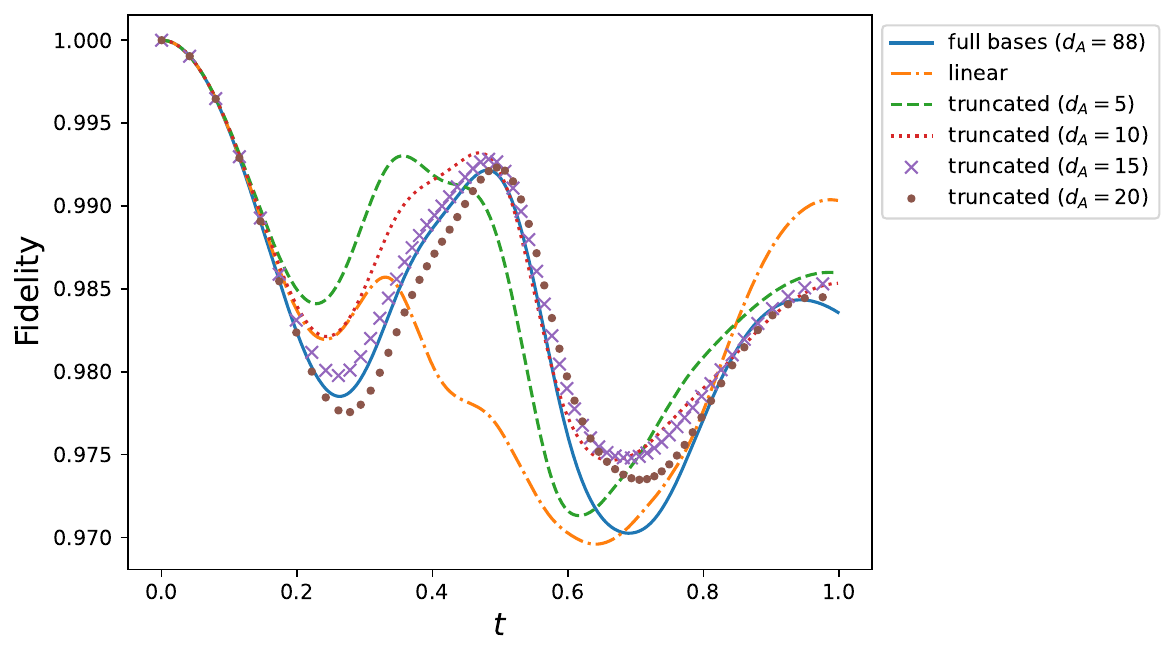}
  \caption{Fidelity of the obtained state to the ground state of the ANNNI model in the antiferromagnetic-like phase at each time $t$. 
  The annealing time $T$ is fixed as 5.91. 
  In several regimes, the fidelity recovers owing to non-adiabatic transitions. 
  Due to such a recovery, the linear schedule eventually shows higher fidelity than our scheme.
  }
  \label{fig: fidelity_vs_t591}
\end{figure}

Figure~\ref{fig: fidelity_vs_t591} shows the fidelity for the annealing time $T=5.91$.
In several regimes, the fidelity recovers and increases because the quantum state $|\Psi(t)\rangle$ returns to the ground state $|\phi_{\mathrm{g}}(\lambda(t))\rangle$ owing to non-adiabatic transitions.
That is, the existence of nonadiabatic transitions plays an important role to achieve the high fidelity (the low relative error) in the adiabatic regime of the present model. 
Since our method is designed for suppressing nonadiabatic transitions, our method cannot incorporate such nonadiabatic improvement.

Finally, we mention possibility of applying our method to large systems. 
In Fig.~\ref{Fig.isingchainene}, we considered the same truncation numbers for different system sizes of the transverse-field Ising chain, i.e., $d_A=1,3,9$ for $L=10$ and $L=50$, and observed significant improvement even for $L=50$. 
We also observed that, with the ANNNI model, obtained schedules and enhanced performance by truncated adiabatic gauge potential are similar to those by the exact adiabatic gauge potential. 
Thus, we believe that our method works even for large systems. 
We could also incorporate optimization or learning methods to further improve schedules obtained by our method. 
We leave further analysis and improvement as the future work.

\section{\label{Sec.conclusion}Conclusion}

We proposed the quantum adiabatic brachistochrone protocol which does not use direct information on energy spectra. 
In this method, we compute the approximate adiabatic gauge potential and utilize the amplitude of it as a measure of nonadiabaticity. 
In particular, the approximate adiabatic gauge potential is calculated by using the variational approach based on the Krylov basis. 
Our method enables us to obtain parameter schedules with classical computers before implementing adiabatic control on quantum devices. 
We believe that our method can easily be realized in current devices and improves performance of quantum annealing.

\begin{acknowledgments}
This work was supported by JST Moonshot R\&D Grant Number JPMJMS2061.
\end{acknowledgments}

\appendix
\section*{Appendix}

Throughout the Appendix, we consider multi-dimensional time-dependent parameter $\bm{\lambda}=(\lambda^1,\lambda^2,\dots)$ for generalization and extension of the present results.

\section{\label{Sec.QAB}Quantum adiabatic brachistochrone}

In quantum adiabatic brachistochrone~\cite{Rezakhani2009,Rezakhani2010}, we introduce an action
\begin{equation}
\epsilon[\bm{\lambda}]=\int\sqrt{2g_{ij}(\bm{\lambda})\dot{\lambda}^i\dot{\lambda}^j}dt,
\label{Eq.action}
\end{equation}
where $g_{ij}(\bm{\lambda})$ is a metric reflecting nonadiabaticity. 
Here, we adopt the Einstein notation of the summation. 
By applying the variational principle to the action, we obtain the geodesic equation
\begin{equation}
\begin{aligned}
&\ddot{\lambda}^i+\Gamma^i_{jk}(\bm{\lambda})\dot{\lambda}^j\dot{\lambda}^k=0, \\
&\Gamma^i_{jk}(\bm{\lambda})=\frac{1}{2}g^{il}(\bm{\lambda})\bm{(}\partial_{k}g_{lj}(\bm{\lambda})+\partial_jg_{lk}(\bm{\lambda})-\partial_lg_{jk}(\bm{\lambda})\bm{)}, 
\end{aligned}
\label{Eq.geodesic.multi}
\end{equation}
where $\partial_l=\partial/\partial\lambda^l$. 
By solving Eq.~(\ref{Eq.geodesic.multi}), we can obtain a geometrically optimal schedule of adiabatic control. 
For a single parameter case, the geodesic equation (\ref{Eq.geodesic.multi}) results in Eq.~(\ref{Eq.geodesic}).

\section{\label{Sec.nonadiabaticity}Counterdiabatic driving}

We consider dynamics $|\Psi(t)\rangle$ governed by the Schr\"odinger equation under a time-dependent Hamiltonian $\hat{H}(\bm{\lambda})$ in the local reference frame. 
By moving to a rotating frame as $|\tilde{\Psi}(t)\rangle=\hat{V}^\dag(\bm{\lambda})|\Psi(t)\rangle$ with a unitary operator $\hat{V}(\bm{\lambda})$, in which the Hamiltonian $\hat{H}(\bm{\lambda})$ is diagonalized, we obtain the Schr\"odinger equation spanned by the energy-eigenstate basis
\begin{equation}
    i\frac{\partial}{\partial t}|\tilde{\Psi}(t)\rangle=\left[\hat{V}^\dag(\bm{\lambda})\hat{H}(\bm{\lambda})\hat{V}(\bm{\lambda})-i\hat{V}^\dag(\bm{\lambda})\bm{(}\partial_t\hat{V}(\bm{\lambda})\bm{)}\right]|\tilde{\Psi}(t)\rangle.
    \label{Eq.Seq.ad}
\end{equation}
Since the first term is diagonalized in the energy-eigenstate basis, the off-diagonal elements of the second term cause nonadiabatic transitions. 
The operator form of the off-diagonal elements of the second term is given by the counterdiabatic Hamiltonian (\ref{Eq.cdham}) with the opposite sign. 
That is, we can eliminate the off-diagonal elements of the second term by applying the counterdiabatic Hamiltonian (\ref{Eq.cdham}) to the original Hamiltonian $\hat{H}(\bm{\lambda})$. 
This is the idea of counterdiabatic driving.

\section{\label{Sec.AGP}Approximate adiabatic gauge potential}

The exact adiabatic gauge potential (\ref{Eq.cdham}) can be rewritten as~\cite{Claeys2019}
\begin{widetext}
\begin{equation}
\hat{\mathcal{A}}(\bm{\lambda})=-\lim_{\epsilon\to+0}\frac{1}{2}\int_{-\infty}^\infty ds\ \mathrm{sgn}(s)e^{-\epsilon|s|}e^{i\hat{H}(\bm{\lambda})s}\bm{(}\bm{\nabla}\hat{H}(\bm{\lambda})\bm{)}e^{-i\hat{H}(\bm{\lambda})s}.
\label{Eq.AGP.integral}
\end{equation}
\end{widetext}
This expression involves fictitious time evolution of $\bm{(}\bm{\nabla}\hat{H}(\bm{\lambda})\bm{)}$ by the Hamiltonian $\hat{H}(\bm{\lambda})$ with the fixed parameter $\bm{\lambda}$, and thus the adiabatic gauge potential is spanned by the Krylov subspace $\{\mathcal{L}^n\bm{\nabla}\hat{H}(\bm{\lambda})\}$. 
More precisely, the adiabatic gauge potential is spanned by the odd Krylov subspace $\{\mathcal{L}^{2n-1}\bm{\nabla}\hat{H}(\bm{\lambda})\}$ as a result of the integral in Eq.~(\ref{Eq.AGP.integral}). 
Similarly, the odd basis operators generated by the Lanczos method (\ref{Eq.Lanczos}) spans the adiabatic gauge potential~\cite{Takahashi2024,Bhattacharjee2023}.

The adiabatic gauge potential satisfies the following equation~\cite{Kolodrubetz2017,Sels2017}
\begin{equation}
\label{eq: adiabatic equation}
\mathcal{L}\left(\bm{\nabla}\hat{H}(\bm{\lambda})-i\mathcal{L}\hat{\mathcal{A}}(\bm{\lambda})\right)=0. 
\end{equation}
This equation can be reformulated as a minimization problem of an action
\begin{equation}
\begin{aligned}
&S[\hat{\mathcal{A}}^\ast(\bm{\lambda})]=\|\hat{G}[\hat{\mathcal{A}}^\ast(\bm{\lambda})]\|_\mathrm{HS}^2, \\
&\hat{G}[\hat{\mathcal{A}}^\ast(\bm{\lambda})]=\bm{\nabla}\hat{H}(\bm{\lambda})-i\mathcal{L}\hat{\mathcal{A}}^\ast(\bm{\lambda}),
\end{aligned}
\end{equation}
regarding a trial adiabatic gauge potential $\hat{\mathcal{A}}^\ast(\bm{\lambda})$. 
We find its argument of the minimum by variational operation
\begin{equation}
\frac{\delta S[\hat{\mathcal{A}}^\ast(\bm{\lambda})]}{\delta\hat{\mathcal{A}}^\ast(\bm{\lambda})}=0,
\label{Eq.variation.action}
\end{equation}
and thus determine $\hat{\mathcal{A}}^\ast(\bm{\lambda})$. 
When the operator form of the trial adiabatic gauge potential $\hat{\mathcal{A}}^\ast(\bm{\lambda})$ is limited, it gives an approximate adiabatic gauge potential.

\section{\label{Sec.geometrical}Geometrical property of the counterdiabatic Hamiltonian}
The rescaled Hilbert-Schmidt norm of the counterdiabatic Hamiltonian is given by 
\begin{equation}
\begin{aligned}
&\|\hat{H}_\mathrm{cd}(t)\|_\mathrm{HS}=\sqrt{\sum_ng^{(n)}_{ij}(\bm{\lambda})\dot{\lambda}^i\dot{\lambda}^j}, \\
&g_{ij}^{(n)}(\bm{\lambda})=\langle\partial_in(\bm{\lambda})|\bm{(}1-|n(\bm{\lambda})\rangle\langle n(\bm{\lambda})|\bm{)}|\partial_jn(\bm{\lambda})\rangle,
\end{aligned}
\label{Eq.strength.cdham}
\end{equation}
where $g_{ij}^{(n)}(\bm{\lambda})$ is the quantum geometric tensor for the $n$th energy eigenstate~\cite{DelCampo2012,Funo2017}. 
Here, we adopt the Einstein notation of the summation. 
The quantum geometric tensor is related to the adiabatic landscape of the energy eigenstate as $|\langle n(\bm{\lambda})|n(\bm{\lambda}+d\bm{\lambda})\rangle|\approx1-(1/2)g_{ij}^{(n)}(\bm{\lambda})d\lambda^id\lambda^j$ and shows a singular behavior at the critical point~\cite{Zanardi2006,Zanardi2007}, where critical slowing down happens. 
That is, the information of the counterdiabatic Hamiltonian (\ref{Eq.cdham}) has ability to find small energy gaps and possibility of nonadiabatic transitions, while the Hilbert-Schmidt norm of the exact counterdiabatic Hamiltonian (\ref{Eq.strength.cdham}) includes information of the all energy eigenstates.


\bibliography{QAschedulingbib}

\end{document}